\documentclass[conference]{IEEEtran}
\ifCLASSINFOpdf
   \usepackage[pdftex]{graphicx}
\else
\fi

\usepackage{amsmath}

\usepackage{algorithmic}
\usepackage{listings}
\makeatletter
\lstdefinelanguage{mlir}{
    classoffset=0,
    morekeywords={
        module,
        func
    },
    morestring=[b]",
    alsoletter={\%},
    keywordsprefix={\%}
}
\makeatother
\usepackage{subcaption}
\usepackage{hyperref}
\hypersetup{
 unicode=true,
 pdfstartview={FitH},
 colorlinks=true,
 citecolor=blue,
 linkcolor=blue,
 urlcolor=blue
}

\usepackage[acronym]{glossaries}
\glsdisablehyper
\newacronym{dfg}{DFG}{dataflow graph}
\newacronym{dsl}{DSL}{domain-specific language}
\newacronym{hbm}{HBM}{high bandwidth memory}
\newacronym{hdl}{HDL}{hardware description language}
\newacronym{hls}{HLS}{high-level synthesis}
\newacronym{ii}{II}{initiation interval}
\newacronym{ir}{IR}{intermediate representation}
\newacronym{mlir}{MLIR}{multi-level intermediate representation}
\newacronym{pc}{PC}{pseudochannel}
\newacronym{plm}{PLM}{private local memory}
\newacronym{tsv}{TSV}{through-silicon via}

\usepackage[table,dvipsnames]{xcolor}
\definecolor{wine}{rgb}{0.6,0.0,0.6}
\definecolor{brick}{rgb}{0.71, 0.2, 0.11}
\definecolor{periwinkle}{rgb}{0.55,0.55,1.0}

\newcommand{\figscale}{0.33}

\begin{document}
%
\title{Platform-Aware FPGA System Architecture Generation based on MLIR}

\author{\IEEEauthorblockN{Stephanie Soldavini}
\IEEEauthorblockA{Dipartimento di Elettronica, Informazione e Bioingegneria\\
Politecnico di Milano, Italy\\
stephanie.soldavini@polimi.it}
\and
\IEEEauthorblockN{Christian Pilato}
\IEEEauthorblockA{Dipartimento di Elettronica, Informazione e Bioingegneria\\
Politecnico di Milano, Italy\\
christian.pilato@polimi.it}
}

\maketitle

\begin{abstract}
FPGA acceleration is becoming increasingly important to meet the performance demands of modern computing, particularly in big data or machine learning applications. As such, significant effort is being put into the optimization of the hardware accelerators. However, integrating accelerators into modern FPGA platforms, with key features such as \gls{hbm}, requires manual effort from a platform expert for every new application. We propose the Olympus \gls{mlir} dialect and Olympus-opt, a series of analysis and transformation passes on this dialect, for representing and optimizing platform aware system level FPGA architectures. By leveraging \gls{mlir}, our automation will be extensible and reusable both between many sources of input and many platform-specific back-ends. 
\end{abstract}

\IEEEpeerreviewmaketitle

\section{Introduction}\glsresetall

As FPGA acceleration is becoming increasingly important, particularly in machine learning and big data applications, significant effort is being put into optimizing the accelerators themselves. Unfortunately, many such applications experience issues with extreme memory bottlenecks \cite{everest}. To overcome this, modern FPGA platforms such as the Xilinx Alveo or Intel Stratix 10, feature \gls{hbm} with many channels to achieve a maximum throughput of over 400GB/s. Using this bandwidth effectively, however, requires very careful handcrafting of system architectures to handle the data movements efficiently. As such, integrating these highly-optimized accelerators into an efficient system leveraging \gls{hbm} requires manual effort from a platform expert for every new application.
The \gls{mlir} framework \cite{mlir} can help make efforts to automate this extensible to the many platforms that exist. 

We propose a toolflow to automatically generate FPGA system architectures optimized for memory bandwidth efficiency leveraging \gls{mlir}. The toolflow consists of the Olympus \gls{mlir} dialect and Olympus-opt, a series of analysis and transformation passes on this dialect, for representing and optimizing platform aware system level FPGA architectures.

\section{Background}
\subsection{MLIR}
\gls{mlir} \cite{mlir} is a novel compiler infrastructure centered on reuse and extensibility. It is becoming popular as a framework for \gls{dsl} compilers for heterogeneous systems, particularly for machine learning. \gls{mlir} is not a single \gls{ir}, but a collection of \emph{dialects}, each representing different layers of abstraction through various operators, types, and attributes. Custom dialects can easily be added to facilitate domain-specific problems while reusing any applicable existing portions of the infrastructure. These dialects can be integrated into larger language stacks via \emph{lowering}. Lowering transforms a more abstract dialect into a more concrete one.

\subsection{FPGA Memory Architecture}
FPGA platforms typically feature multiple kinds of memory. First, there are BRAM or URAM memory elements, which can be aggregated to form \gls{plm}. Then, FPGAs are often integrated with DRAM. A common DRAM technology used with FPGAs today is DDR4, which has a 64-bit data interface for each module. Typical systems have two modules and so two \emph{channels} for a total bitwidth of 128 bits. For increased bandwidth, modern FPGA platforms such as the Xilinx Alveo U280 and the Intel Stratix 10 MX are integrated with \gls{hbm}~\cite{Choi2020}.
\gls{hbm} is a 3D-stacked, DRAM-based memory architecture, exposing many parallel channels to the FPGA logic and allowing for high-bandwidth and energy-efficient data movements~\cite{Fujita2021}.

This work uses the \textbf{Xilinx Alveo U280} data center accelerator card as an example target platform, but other devices would benefit from the same system-level optimizations.
The Alveo U280 features the XCU280 FPGA, built on the Xilinx 16nm UltraScale+ architecture and offers both DDR4 and \gls{hbm}. The U280 has 2 DDR4 banks of 16~GB each for a total DDR bandwidth of 38~GB/s. The U280 interfaces with the \gls{hbm}2 subsystem through 32 \textbf{\glspl{pc}} each directly accessing a 256~MB memory bank (8~GB in total). Each 256-bit \gls{pc} operates at 450~MHz, for a maximum bandwidth of 14.4~GB/s. Therefore the theoretical maximum bandwidth of the full \gls{hbm} is 460.8~GB/s. 

\section{Related Work}

While \gls{mlir} was designed primarily for software compilers, many concepts can also be applied to hardware design tools. A few works use \gls{mlir} as the basis for their tools.
SODA-OPT~\cite{Agostini2022} is a compiler tool extending the \gls{mlir} infrastructure to generate FPGA accelerators through \gls{hls}. SODA-OPT can generate an FPGA or ASIC design and host executable to implement the overall input program.
ScaleHLS~\cite{Ye2022} is an \gls{hls} framework built with \gls{mlir} to optimize accelerators at multiple levels of representation. ScaleHLS provides multiple analyses and transformation passes and a DSE engine to optimize designs automatically.
HECTOR \cite{Xu2022} is a two-level \gls{ir} built using the \gls{mlir} framework for representing hardware accelerators and converting them into RTL designs. The ``ToR'' \gls{ir} is higher level and software-like with temporal representation, while the ``HEC'' \gls{ir} is lower level with a more spatial representation.
CIRCT \cite{circt} attempts to extend \gls{mlir} to hardware design and acts as the hardware \gls{ir} whereas most synthesis tools today use VHDL or Verilog as \gls{ir}, both of which cannot benefit from any of more abstract design characteristics. 
These tools are all focused on optimizing individual accelerators, or a whole system within an FPGA, but there is no focus on optimizing the global memory access and bandwidth bottleneck. 

Additionally, some hardware design automation works mention integrating with \gls{mlir} as a helpful ``future work.'' 
The work in \cite{Forget2022} presents a high-level optimization framework, demonstrated with a C++20 library for fixed-point arithmetic and a compiler flow from C++20 to Vivado HLS. They mention that \gls{mlir} may be useful for future work instead of relying on custom compiler components. 
PipeArch \cite{Kara2021} is a tool combining the efficiency of specialized hardware accelerators with the generality of CPU threads. Portions of the design were manual, however, and they have suggested future work would be automation leveraging the \gls{mlir} framework.
\gls{mlir} thus shows its utility as a common interface between tools, promoting reusability.

\section{Olympus Dialect}

The Olympus Dialect is designed to represent the \gls{dfg} of kernels to be offloaded to FPGA. The \gls{dfg} is composed of two operators representing kernels (nodes) and channels (edges). 
A sample operator for making a channel is shown in \autoref{fig:channelop}.
\begin{figure}[ht!]
    \centering
    \begin{lstlisting}
%2 = "olympus.make_channel"() {
    encapsulatedType = i32, 
    paramType = "stream",
    depth = 20
} : () -> (
    !olympus.channel<i32>
)
    \end{lstlisting}
    \caption{Sample channel operator}
    \label{fig:channelop}
\end{figure}

The attributes of the \texttt{olympus.make\_channel} operator are \texttt{encapsulatedType}, \texttt{paramType}, \texttt{depth}. The \texttt{encapsulatedType} is a signless integer of arbitrary bitwidth. The interpretation of the data is not important, only the width. Therefore a 32-bit float, a fixed-point value with 10 integer bits and 22 fraction bits, and a 32-bit integer should all be represented as `\texttt{i32}'.
The \texttt{paramType} describes the properties of the data in one of three ways: \texttt{stream}, \texttt{small}, or \texttt{complex}. ``\texttt{stream}'' data must be produced and consumed in the same order and consist of small, statically sized elements. ``\texttt{small}'' data can be random access, but in total the data needed for a single kernel iteration should be at most on the scale of 100s of kB and be organized of simple structures without nesting or indirection. ``\texttt{complex}'' data can be anything: huge, random access, have indirection, and/or be constructed of nested structures. The \texttt{depth} attribute describes how large the data is in total. If \texttt{paramType==stream}, \texttt{depth} is the maximum necessary channel depth. If \texttt{paramType==small}, \texttt{depth} is the number of elements. If \texttt{paramType==complex}, \texttt{depth} is the number of bytes. The return value is a \texttt{olympus.channel} type with the \texttt{encapsulatedType} as the element type. The return value is used as an operand to kernel operators to represent the channel connections. 

\begin{figure}[ht]
    \centering
    \begin{lstlisting}
"olympus.kernel"(%2, %3, %4) {
    callee = "matmul", 
    latency = 795, ii = 268,
    ff = 3106, lut = 6174, bram = 61, 
    uram = 0, dsp = 48,
    operand_segment_sizes = array<i32: 2, 1>,
} : (
    !olympus.channel<i32>, 
    !olympus.channel<i32>, 
    !olympus.channel<i32>
) -> ()
\end{lstlisting}
    \caption{Sample kernel operator}
    \label{fig:kernelop}
\end{figure}

A sample kernel operator is shown in \autoref{fig:kernelop}.
The attributes of the \texttt{olympus.kernel} operator are \texttt{callee}, \texttt{latency}, \texttt{ii}, and one attribute for each FPGA resource quantity. The \texttt{callee} attribute is the name of the kernel function that this kernel should execute. This is used to find the correct implementation of the kernel when generating the hardware. 
The \texttt{latency}, \texttt{ii} (\glsentrylong{ii}), and resources (\texttt{ff, lut, bram, uram, dsp}) attributes are the timing and resource estimates.
Additionally, there is an \texttt{operand\_segment\_sizes} attribute to delineate which of the operands are inputs and which are outputs. In this case, the first two operands (\texttt{\%2, \%3}) are inputs and the last operand (\texttt{\%4}) is an output. In this way, multiple outputs are allowed.

This \gls{ir} can be lowered from a higher level \gls{mlir} dialect, particularly one focused on the \gls{dfg} flow of an application, or generated by a \gls{dsl} compiler focusing on domains which benefit from FPGA acceleration. 

\section{Olympus Lowering}
\begin{figure}[ht]
    \centering
    \includegraphics[width=\columnwidth]{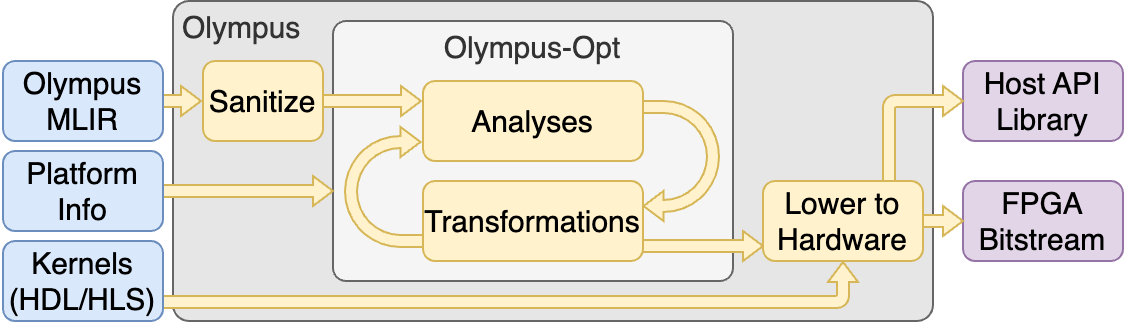}
    \caption{Olympus flow diagram: starting from an \gls{mlir} system level description, platform info, and kernel implementations Olympus generates an optimized hardware architecture implemented as an FPGA bitstream and host API library.}
    \label{fig:flow}
\end{figure}

From its \gls{mlir} system-level description, Olympus optimizes and generates a hardware system architecture using the xDSL \cite{xdsl} library to perform transformations on \gls{mlir} using Python. A diagram of the overall flow is shown in \autoref{fig:flow}. The inputs to Olympus, shown on the left in blue are the Olympus \gls{mlir} description of the \gls{dfg}, the FPGA platform details, and the kernel implementations. The kernel can be in the form of Vitis HLS, HDL generated by other HLS tools (such as Bambu \cite{bambu}), or custom HDL. Olympus performs sanitation of the input, then iterates over the Olympus-Opt analyses and transformations to optimize the final \gls{dfg}. Finally, the \gls{dfg} is lowered to hardware and the output products, shown in purple on the right, are produced for both the host driver API library and the FPGA bitstream.

\subsection{Sanitize step}
\begin{figure}[!t]
\centering
    \begin{subfigure}[b]{0.6\columnwidth}
    \begin{subfigure}{\columnwidth}
       \centering
       \includegraphics[scale=\figscale]{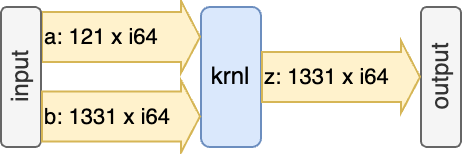}
       \caption{Original input \gls{dfg}}
       \label{fig:input}
    \end{subfigure}
    
    \begin{subfigure}{\columnwidth}
       \centering
       \includegraphics[scale=\figscale]{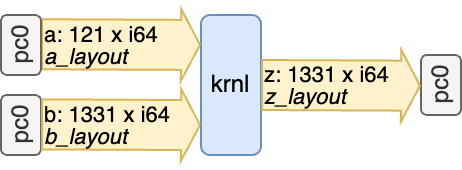}
       \caption{{Sanitized DFG}}
       \label{fig:san}
    \end{subfigure}
    \end{subfigure}
    \hfill
    \begin{subfigure}[b]{0.38\columnwidth}
       \centering
       \includegraphics[scale=\figscale]{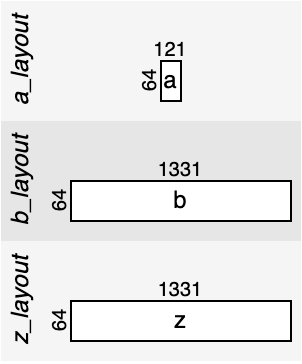}
       \caption{Sanitized input layouts}
       \label{fig:san_layout}
    \end{subfigure}
\caption{Visualization of a \gls{dfg} with one kernel with two input and one output channel. The original input \subref{fig:input} is sanitized to be \subref{fig:san} with \gls{pc} nodes and layouts \subref{fig:san_layout} for each channel.}
\label{fig:input_to_san}
\end{figure}

The first step is to sanitize the input Olympus \gls{mlir}, visualized in \autoref{fig:input}, into a form that could immediately be passed to the hardware lowering step to create the system architecture. This allows the user to create the \gls{mlir} in a more convenient form without having to add redundant details.

First, layouts are created for each channel. The layout is an additional attribute of the channel operators and represents the organization of the data when sent through the channel. The layout created at this stage is simply a width of one element and a depth of the \texttt{depth} attribute, shown in \autoref{fig:san_layout}. 

Additionally, \texttt{olympus.pc} nodes are created for each data channel connected to global memory (i.e. not connected to kernels on both sides). These are similar to kernel operations but instead represent the \gls{pc} of global memory and are used as the terminals for data channels to main memory. These operations have one attribute (the \texttt{id} of the memory channel) and one operand (the channel connected to this \gls{pc}). The direction is inferred by whether this channel is an input or output for the kernel it is connected to. In this stage, each channel to global memory is connected to one \texttt{olympus.pc} node and all \texttt{id} attributes are set to 0.

After these steps, the \gls{ir} can be immediately lowered to \gls{hdl} and synthesized into a working, but inefficient, design (\autoref{fig:san}).

\subsection{Olympus-Opt}
The next stage is an iterative series of analyses and transformations to obtain a more optimized system architecture. In addition to the sanitized input \gls{mlir}, this stage requires the FPGA target specification including: the number of global memory channels and their widths and the amounts of each available resource. Additionally, a resource utilization limit (default 80\%) can be given.

The analyses comprise of two main calculations. First, the target \gls{pc} information and the attributes of each data channel are used to calculate a bandwidth utilization percentage. Second, the total resource availability and the kernel resource utilization are used to estimate an overall utilization.

\begin{figure}[ht]
   \centering
   \includegraphics[scale=\figscale]{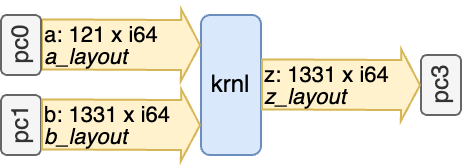}
   \caption{Sample result of applying channel reassignment to \autoref{fig:san}. Each PC node has been given a different \texttt{id}.}
   \label{fig:chan}
\end{figure}

Using the results of these analyses, transformation passes can be chosen to alter the \gls{dfg} to increase expected performance. These transformations include the following:

\textbf{Channel reassignment:} Data channels connected to \gls{pc} nodes and data channels of \texttt{complex} type are distributed across the channels available on device to increase bandwidth utilization. \autoref{fig:chan} shows how \autoref{fig:san} would be transformed with each \gls{pc} node being assigned a separate \texttt{id} number, to represent a mapping onto separate physical \glspl{pc}. 

\begin{figure}[ht]
   \centering
   \includegraphics[scale=\figscale]{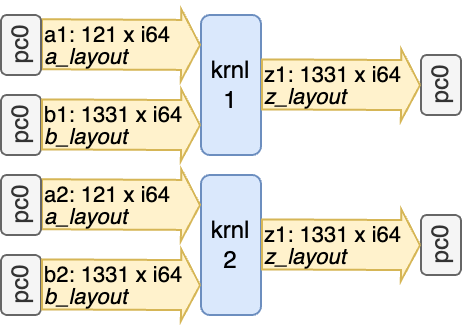}
   \caption{Sample result of replicating \autoref{fig:san} two times.}
   \label{fig:repl}
\end{figure}

\textbf{Replication:} If the resource utilization is low, the entire \gls{dfg} can be replicated for increased parallelism, up to the resource utilization limit. \autoref{fig:repl} shows how \autoref{fig:san} would be replicated twice. 
Each operator is replicated and given a new identifier. Each replicated \gls{pc} node is given the same \texttt{id}. Replication can gain near ideal speedup, however a high degree of replication reaching near 100\% utilization of a resource induces routing congestion and therefore a longer critical path. Replication should be used carefully, utilizing other optimizations for more performance.

\begin{figure}[ht]
    \captionsetup[subfigure]{skip=-37pt,slc=off,margin={-3pt,0pt}}
    \begin{subfigure}[b]{\columnwidth}
       \centering
       \includegraphics[scale=\figscale]{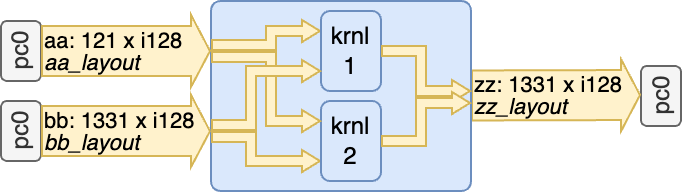}
       \caption{}
       \label{fig:wide_sub}
    \end{subfigure}
    
    \vspace{30pt}
    \captionsetup[subfigure]{skip=-28pt,slc=off,margin={-3pt,0pt}}
    \begin{subfigure}[b]{\columnwidth}
       \centering
       \includegraphics[scale=\figscale]{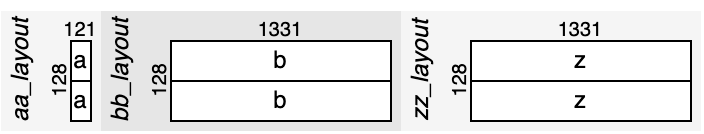}
       \caption{}
       \label{fig:wide_layout}
    \end{subfigure}
    \vspace{10pt}
    \caption{Sample result of applying bus widening to \autoref{fig:san} with a bus width of 128. Each channel has been widened by $2\times$, and two kernels are instantiated. The layouts \subref{fig:wide_layout} have each data array replicated in parallel.}
    \label{fig:wide}
\end{figure}

\textbf{Bus widening:} If data widths are evenly divisible into \gls{pc} widths, kernels can be replicated such that multiple instances use the full \gls{pc}. For instance, a kernel with a 64-bit data input using a 256-bit \gls{pc} can be replicated four times so each kernel's data uses one of four lanes in the \gls{pc}~\cite{TRETS22}. \autoref{fig:wide} shows how \autoref{fig:san} would be affected by bus widening for a 128-bit bus. 
Each data channel is made twice as wide and the layout is modified to act as two ``lanes''. These channels are connected to a super-node encapsulating two kernels. When this is lowered to hardware, the data mover modules separate the ``lanes'' and send the data to the correct kernels. With sufficient resource availability, this optimization achieves near ideal speedup for the number of replications.

\begin{figure}[t]
    \captionsetup[subfigure]{skip=-30pt,slc=off,margin={30pt,0pt}}
    \begin{subfigure}[b]{\columnwidth}
       \centering
       \includegraphics[scale=\figscale]{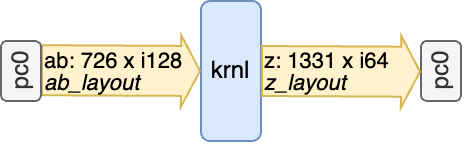}
       \caption{}
       \label{fig:iris_sub}
    \end{subfigure}
    
    \vspace{23pt}
    \captionsetup[subfigure]{skip=-28pt,slc=off,margin={30pt,0pt}}
    \begin{subfigure}[b]{\columnwidth}
       \centering
       \includegraphics[scale=\figscale]{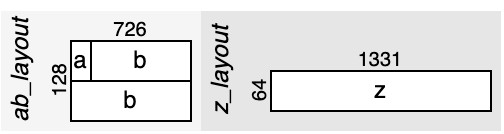}
       \caption{}
       \label{fig:iris_layout}
    \end{subfigure}
    \vspace{10pt}
    \caption{Sample result of applying the Iris algorithm to \autoref{fig:san} to combine the $a$ and $b$ channels on a 128-bit bus. $a$ and $b$ are interleaved in the layout \subref{fig:iris_layout} of the new $ab$ channel.}
    \label{fig:iris}
\end{figure}

\textbf{Bus optimization:} To increase bandwidth utilization, channels can be grouped to interleave data~\cite{ASPDAC23}. The Iris algorithm can split data into smaller chunks and interleave them with other arrays to compact them on a bus with a given width. \autoref{fig:iris} shows how Iris combines the $a$ and $b$ channels in  \autoref{fig:san} into a 128-bit bus. In the new single channel, the layout reflects the result of the Iris algorithm with the $b$ array broken up to achieve the most compact result. The Iris algorithm can achieve over 95\% bandwidth efficiency for a channel, compared with \~45\% efficiency of a naive layout.

\textbf{PLM optimization:} If the characteristics of the data accesses are known, the physical memories can be shared for area efficiency~\cite{Pilato2017}. Memories or interfaces can be shared based on spatial or temporal compatibility, respectively. This information can be detected by static compiler analysis and supplied as additional information to enable this optimization. This optimization saves on hardware resources, often to a high enough degree to allow for additional compute unit replication and therefore speedup.

\subsection{Lower to Hardware}

After the Olympus-opt passes, a hardware system architecture can be generated. We use Xilinx Alveo platforms as an example, but other back ends can be implemented if they provide implementation adhering to the following description. 

Channels connected to \texttt{olympus.pc} nodes are connected to the \glspl{pc} on the device. For the Alveos, this is configured in the \texttt{*.cfg} file input to the Vitis tool.

Data channels with the \texttt{stream} type are instantiated as FIFOs of the specified \texttt{depth}. \texttt{small} type channels are instantiated as \gls{plm} in BRAMs so data can be randomly accessed, but does not need to be sent out to global memory. These memories can be shared using Mnemosyne-generated \gls{plm} architectures. \texttt{complex} type channels are connected to the device \glspl{pc} so the kernels can use arbitrary pointers to access this data.
Channels with Iris-generated layouts are instantiated with adapters generated by the Iris tool to pack or unpack the data in a way the kernels can use. 

For Xilinx devices, these modules are connected in a Vivado block diagram. One Vitis HLS module is instantiated alongside the kernels to bridge the global memory and the kernels and includes the \glspl{plm} and data moving modules. If a kernel is connected to a \texttt{complex} channel, this kernel has an AXI port that connects directly to the global memory.

Additionally, Olympus generates a host API library for initializing the device, creating on-device data buffers, moving data between host and device memory, and initiating kernel execution. For the Alveo, these functions call the OpenCL Xilinx runtime methods. Other back-ends can implement the same host API using the platform-specific underlying methods.

\section{Conclusion}
We proposed the Olympus MLIR-based infrastructure for platform-aware FPGA system architecture generation. The Olympus MLIR dialect is designed to represent the \gls{dfg} of accelerator kernels. Olympus-opt is a collection of analyses and transformations on this \gls{dfg} to iteratively optimize the \gls{dfg} to take advantage of the characteristics of the FPGA platform, particularly off chip memory bandwidth. 

\section*{Acknowledgment}
This work was partially funded by the EU Horizon 2020 Programme under grant agreement No 957269 (EVEREST).


\bibliographystyle{IEEEtran}
\bibliography{main}
%
\end{document}